\documentstyle[mncite]{mn}
\newif\ifAMStwofonts

\ifoldfss
  \ifCUPmtlplainloaded \else
    \NewTextAlphabet{textbfit} {cmbxti10} {}
    \NewTextAlphabet{textbfss} {cmssbx10} {}
    \NewMathAlphabet{mathbfit} {cmbxti10} {} 
    \NewMathAlphabet{mathbfss} {cmssbx10} {} 
  \fi
  \ifAMStwofonts
    \ifCUPmtlplainloaded \else
      \NewSymbolFont{upmath} {eurm10}
      \NewSymbolFont{AMSa} {msam10}
      \NewMathSymbol{\upi}     {0}{upmath}{19}
      \NewMathSymbol{\umu}     {0}{upmath}{16}
      \NewMathSymbol{\upartial}{0}{upmath}{40}
      \NewMathSymbol{\leqslant}{3}{AMSa}{36}
      \NewMathSymbol{\geqslant}{3}{AMSa}{3E}

      \let\leq=\leqslant 
      \let\geq=\geqslant 
    \fi
  \fi
\fi 

\ifnfssone
  \newmathalphabet{\mathit}
  \addtoversion{normal}{\mathit}{cmr}{m}{it}
  \addtoversion{bold}{\mathit}{cmr}{bx}{it}
  \newmathalphabet{\mathbfit} 
  \addtoversion{normal}{\mathbfit}{cmr}{bx}{it}
  \addtoversion{bold}{\mathbfit}{cmr}{bx}{it}
  \newmathalphabet{\mathbfss} 
  \addtoversion{normal}{\mathbfss}{cmss}{bx}{n}
  \addtoversion{bold}{\mathbfss}{cmss}{bx}{n}
  \ifAMStwofonts
    \ifCUPmtlplainloaded \else
      \UseAMStwoboldmath
      \makeatletter
      \new@mathgroup\upmath@group
      \define@mathgroup\mv@normal\upmath@group{eur}{m}{n}
      \define@mathgroup\mv@bold\upmath@group{eur}{b}{n}
      \edef\UPM{\hexnumber\upmath@group}
      \new@mathgroup\amsa@group
      \define@mathgroup\mv@normal\amsa@group{msa}{m}{n}
      \define@mathgroup\mv@bold\amsa@group{msa}{m}{n}
      \edef\AMSa{\hexnumber\amsa@group}
      \makeatother
      \mathchardef\upi="0\UPM19
      \mathchardef\umu="0\UPM16
      \mathchardef\upartial="0\UPM40
      \mathchardef\leqslant="3\AMSa36
      \mathchardef\geqslant="3\AMSa3E

      \let\leq=\leqslant 
      \let\geq=\geqslant 
    \fi
  \fi
\fi 

\ifnfsstwo
  \DeclareMathAlphabet{\mathbfit}{OT1}{cmr}{bx}{it}
  \SetMathAlphabet\mathbfit{bold}{OT1}{cmr}{bx}{it}
  \DeclareMathAlphabet{\mathbfss}{OT1}{cmss}{bx}{n}
  \SetMathAlphabet\mathbfss{bold}{OT1}{cmss}{bx}{n}
  \ifAMStwofonts
    \ifCUPmtlplainloaded \else
      \DeclareSymbolFont{UPM}{U}{eur}{m}{n}
      \SetSymbolFont{UPM}{bold}{U}{eur}{b}{n}
      \DeclareSymbolFont{AMSa}{U}{msa}{m}{n}
      \DeclareMathSymbol{\upi}{0}{UPM}{"19}
      \DeclareMathSymbol{\umu}{0}{UPM}{"16}
      \DeclareMathSymbol{\upartial}{0}{UPM}{"40}
      \DeclareMathSymbol{\leqslant}{3}{AMSa}{"36}
      \DeclareMathSymbol{\geqslant}{3}{AMSa}{"3E}

      \let\leq=\leqslant 
      \let\geq=\geqslant 
    \fi
  \fi
\fi 

\ifCUPmtlplainloaded \else
  \ifAMStwofonts \else 
    \def\upi{\pi}
    \def\umu{\mu}
    \def\upartial{\partial}
  \fi
\fi

\title{The mass and dynamical state of Abell 2218}
\author[D. B. Cannon, T. J. Ponman \& I. S. Hobbs]
       {D. B. Cannon, T. J. Ponman \& I. S. Hobbs\\
        School of Physics and Space Research, University of Birmingham,
        Birmingham B15 2TT, UK \\}
\date{Accepted 1998 ??.
      Received 1998 ??;
      in original form 1997 ??}

\pagerange{\pageref{firstpage}--\pageref{lastpage}}
\def\LaTeX{L\kern-.36em\raise.3ex\hbox{a}\kern-.15em
    T\kern-.1667em\lower.7ex\hbox{E}\kern-.125emX}
\pubyear{1998}

\newcommand{\gtapp}{\mbox{\raisebox{-1mm}{$\stackrel{>}{\sim}$}}}

\begin{document}


\maketitle

\label{firstpage}

\begin{abstract}
Abell 2218 is one of a handful of clusters in which X-ray and lensing
analyses of the cluster mass are in strong disagreement. It is also a
system for which X-ray data and radio measurements of the
Sunyaev-Zel'dovich decrement have been combined in an attempt to
constrain the Hubble constant. However, in the absence of reliable
information on the temperature structure of the intracluster gas, most
analyses have been carried out under the assumption of isothermality. We
combine X-ray data from the ROSAT PSPC and the ASCA GIS instruments,
enabling us to fit non-isothermal models, and investigate the impact that
this has on the X-ray derived mass and the predicted Sunyaev-Zel'dovich
effect.

We find that a strongly non-isothermal model for the intracluster gas,
which implies a central cusp in the cluster mass distribution, is
consistent with the available X-ray data and compatible with the lensing
results. At r$<1\arcmin$, there is strong evidence to suggest that the
cluster departs from a simple relaxed model. We analyse the dynamics of
the galaxies and find that the central galaxy velocity dispersion is too
high to allow a physical solution for the galaxy orbits. The quality of
the radio and X-ray data do not at present allow very restrictive
constraints to be placed on H$_{0}$. It is apparent that earlier analyses
have under-estimated the uncertainties involved. However, values greater
than $50$ km s$^{-1}$ Mpc$^{-1}$ are preferred when lensing constraints
are taken into account.
\end{abstract}


\section{Introduction}

The masses of galaxy clusters can be determined in three main ways: using
the velocity dispersion of the galaxies, the pressure gradient in the hot
intracluster gas derived from X-ray imaging and spectroscopy, and by
analysing the lensing of background galaxies by the cluster potential.
Each of these approaches involve assumptions and are vulnerable to
various systematic errors. It is therefore useful to compare the results
of the different techniques. The X-ray and lensing approaches are
generally considered the most reliable. Results from them have now been
compared for a number of clusters
\cite{fort94a}. The agreement is often reasonable, but there are a few
spectacular exceptions, of which Abell 2218 (hereafter A2218) is the most well 
studied example.

A2218 is an optically compact (core radius $\approx 1 \arcmin$;
\pcite{dress78a}) cluster of galaxies, located at a redshift of 0.171
\cite{krist78a}, and classified as richness class 4 \cite{abell89a}. The
cluster appears well relaxed, with the majority of the galaxies centred
around the sole cD galaxy. However, detailed photometric studies
(\pcite{pello88a}; \pcite{pello92a}) suggest the existence of a second,
smaller galaxy concentration, displaced from the cD by 67$\arcsec$.
Spectroscopic study \cite{lebor92a}, performed on 
the central region ($<4 \arcmin$) of A2218, has
provided redshift information for 66 of the objects within the core and
shown that the average velocity dispersion is 1370 km s$^{-1}$.

A succession of X-ray telescopes have allowed the properties of the hot
gas within A2218 to be established. With the Einstein IPC \& HRI
\cite{perr81a} and ROSAT PSPC \cite{sidd95a}, the emission was found to
be smooth (on scales $\sim 1\arcmin$), azimuthally symmetric and centred
on the cD galaxy. Fitting a polar profile of the surface brightness with
a King model gave a core radius of $58\arcsec^{+16}_{-16}$ and a
$\beta$-value of 0.63 \cite{boyn82a}. Integrated spectral analyses with
Ginga gave a gas temperature of $6.72^{+0.5}_{-0.4}$ keV and metallicity
of $0.2^{+0.2}_{-0.2}$ Z$_{\odot}$ \cite{mchard90a}. By virtue of Ginga's
bandwidth, this determination is commonly accepted as the most accurate
estimate of the mean gas temperature. Most recently, deep observations
with the ROSAT HRI \cite{markev97a} have shown the presence of
significant X-ray substructure within the cluster core, suggesting that
the cluster may have undergone a recent merger event. This may account
for the absence of any signs of a cooling flow in the cluster
(\pcite{arnaud91a}; \pcite{white96a}).

Several previous comparisons (\pcite{miral95a}; \pcite{kneib95a};
\pcite{kneib96a}; \pcite{natar96a}) between strong lensing and X-ray
analyses have found a factor of 2 discrepancy in the gravitating masses
predicted by the two methods. Suggested explanations have centred upon
the assumption of hydrostatic equilibrium for the cluster gas, the
possibility that magnetic fields may provide significant pressure support
to the gas and the presence of substructure within the cluster.

The Sunyaev-Zel'dovich decrement associated with A2218 has been
extensively studied (\pcite{jones93a}; \pcite{birk94a};
\pcite{saund96a}). These results have been used, in conjunction with
X-ray data, to constrain the Hubble constant (\pcite{silk78a};
\pcite{mchard90a}; \pcite{birk94a}). These analyses have been made in the
absence of reliable information about temperature variations in the
intracluster gas and have therefore been forced to make simplifying
assumptions, such as that of isothermality (\pcite{mchard90a};
\pcite{birk94a}; \pcite{kneib95a}). This assumption is without a strong
theoretical foundation and conflicts with the results of most
cosmological simulations (\pcite{navar95a}; \pcite{navar96a};
\pcite{tormen96a}), which show temperature declining with radius, and
mass distributions which have a central cusp. The question then arises as
to whether simplifying assumptions have significantly biased the
conclusions of previous X-ray analyses. For example, is the apparent
discrepancy between the X-ray and lensing masses unavoidable or does it
arise simply from the use of inappropriate assumptions in the X-ray
analysis?

Motivated by the desire to avoid such restrictive assumptions, we have
carried out an X-ray analysis which combines the capabilities of ROSAT
and ASCA. The limited spectral bandwidth and resolution of the ROSAT PSPC
is compensated for by the superior spectral properties of ASCA.
Conversely, the poor spatial performance of ASCA is complemented by the
higher spatial resolution of ROSAT. This approach has never before been
applied to A2218.

The central aim of this paper is to improve our understanding of A2218 by
comparing the results of our X-ray analysis with lensing, SZ and galaxy
velocity studies. It also serves as a case study on the possible dangers
of assuming an isothermal gas, when one has no information to the
contrary. Throughout the paper we assume an Einstein-de Sitter cosmology
with $\Omega$=1, q$_0$=0.5 and H$_0=50$ km s$^{-1}$ Mpc$^{-1}$, except
where otherwise stated.

\section{X-ray Analysis}

The objective of the analysis is to use spatially and spectrally resolved
X-ray data to constrain models of the distribution of gas properties in
the cluster. For an in-depth discussion of the procedures covered in this
section, see \scite{cannon97a}.

\subsection{Spectral-image modelling}

We work with X-ray spectral images, which constitute blurred records of
the spectral properties of the cluster projected along the line of sight.
Since information about the disposition of material perpendicular to the
plane of the sky is not available, it is necessary to make some
assumption about the geometry of the source. We assume that the cluster
is spherically symmetric. In practice, A2218 is slightly elliptical, with
an axis ratio of 0.8 \cite{sidd95a}. 
However this modest ellipticity should not introduce any serious
errors into our derived masses \cite{fabric84a}.

It is important to allow for the spatial and spectral blurring introduced
by the telescope, as described by the instrument point spread function
(psf) and energy response matrix. We adopt a forward fitting approach
(\pcite{eyles91a};
\pcite{watt92a}), in which the properties of the gas are parameterised as
analytical functions of cluster radius. The emission from each spherical
shell in the cluster is computed using a \scite{raym77a}, hereafter RS,
hot plasma code. After correcting for cluster redshift, the spectral
emissivity profiles are folded through the instrument spectral response,
projected along the line-of-sight, rebinned into an $xy$ grid and blurred
with the psf. This produces a predicted spectral image which can be
directly compared to the observed data, using a maximum-likelihood
statistic. Iteratively adjusting the model parameters results in a
best-fit to the data.

Using analytical forms for the radial distribution of gas properties has
the advantage of regularising the solution (i.e. suppressing
instabilities in the deprojection and deblurring processes), however one
runs the risk that the solution may be dictated by the mathematical
function imposed. This can lead to overconfidence in derived results, as
acceptable alternatives which might fit the data are ruled out by the
limitations of the available models. The commonly employed restriction of
isothermality is an extreme example of this. We attempt to avoid this
problem by using a range of radial functions. This is particularly
important for the temperature and we use not only a number of parametric
forms for T$_{gas}(r)$, but also an alternative approach in which
T$_{gas}(r)$ is determined indirectly, by fitting a model for the mass
distribution, as discussed below.  The gas density profile is much more
readily determined by the X-ray data, so we have fitted only two radial
forms.

Assuming that the intracluster gas is in hydrostatic equilibrium in the
potential well of the cluster, the total gravitating mass within radius
r, from the centre of the cluster, is related to the gas temperature and
density by:

\begin{equation}
M_{grav}(r) = - {\frac {kT_{gas}(r)}{G \mu m_{p}}} \left[ {\frac
 {d \ln \rho_{gas}(r)}{d \ln r}} + { \frac {d \ln T_{gas}(r)}{d \ln r}}
 \right] r
\label{eq:xray_hydro}
\end{equation}
where $\rho_{gas}(r)$ is the gas density, T$_{gas}(r)$ the gas temperature, 
$\mu$ the mean molecular weight and $m_{p}$ the proton mass.

\subsubsection{Gas density}

The gas density is well constrained by the X-ray surface brightness,
since the PSPC is largely insensitive to variations in temperature for
T$>3$~keV. Surface brightness profiles are generally well fitted by
core-index type models (\pcite{king62a}; \pcite{king72a}):

\begin{equation}
\rho_{gas}(r)=\rho_{gas,0}\left[1+(r/r_{c})^2\right]^{-\alpha_{\rho}}
\end{equation}
where $\rho_{gas,0}$ is the central gas density normalisation (amu cm$^{-3}$),
$r_{c}$ the core radius (arcmin) and $\alpha_{\rho}$ the density index 
(unitless). The main deviations from this form occur at small radii, where 
cooling flows give rise to surface brightness cusps in many clusters, though 
not in A2218.

Recent N-body studies (\pcite{navar95a}; \pcite{navar96a}; \pcite{tormen96a})
have achieved good fits to dark matter (DM) and gas profiles in simulated 
clusters with an alternative description. The profiles are found to steepen 
progressively, from $\rho_{gas}(r) \propto r^{-1}$ in the core, to $r^{-3}$ 
near the virial radius, following the form

\begin{equation}
\rho_{gas}(r) = \rho_{gas,0} \left[ x (1+x)^2 \right]^{-1}
\end{equation}
where $x = r/r_{s}$, $r_{s}$ being the scale radius (arcmin). Both of the
above analytical forms have been fitted to the X-ray data for A2218.

\subsubsection{Gas temperature}

The gas temperature distribution is less well determined, since this requires
a combination of spatial and spectral resolution which has not generally been 
available in the past. We consider a variety of simple models: a linear 
temperature ramp (LTF),

\begin{equation}
T_{gas}(r) = T_{gas,0} - \beta r
\end{equation}
where $T_{gas,0}$ is the gas temperature (keV) at the cluster centre, $\beta$ 
the temperature gradient (keV arcmin$^{-1}$) and $r$ the radius (arcmin); a 
King-type temperature description (KTF),

\begin{equation}
T_{gas}(r) = T_{gas,0} \left[ 1 + (r/r_{\mbox{\tiny T}})^2 \right]^{-\beta}
\end{equation}
where $r_{\mbox{\tiny T}}$ is the temperature core radius (arcmin) 
and $\beta$ the temperature index (unitless); and a 
polytropic temperature description (TTF),

\begin{equation}
T_{gas}(r) = T_{gas,0} \left[ 1 + (r/r_{c})^2 \right]^{\alpha_{\rho}
({1-\gamma})}
\end{equation}
where $r_{c}$ is the gas density core radius (arcmin) and $\gamma$, the 
polytropic index (unitless), is fitted as a free parameter varying between 
isothermality ($\gamma$=1) and adiabaticity ($\gamma$=5/3).

\subsubsection{Gravitating mass}

An alternative to fitting $\rho_{gas}(r)$ and $T_{gas}(r)$ is to fit
$\rho_{gas}(r)$ and M$_{grav}(r)$. The corresponding temperature profile
can then be inferred, via Equation~\ref{eq:xray_hydro}. We use several
alternative forms, motivated by the distribution of galaxies in clusters
\cite{rood72a}, and by the results of N-body simulations. These include:
a core-index description (DMF),

\begin{equation}
\rho_{\mbox{\tiny DM}}(r)=\rho_{\mbox{\tiny DM,0}}\left[1+(r/r_{c})^2
 \right]^{-\alpha_{\rm DM}}
\label{eq:rho_dm}
\end{equation}
where $\rho_{\mbox{\tiny DM,0}}$ is the central dark matter density 
normalisation (amu cm$^{-3}$), $r_{c}$ the core radius (arcmin) and 
$\alpha_{\rm DM}$ the density index (unitless); a model based upon the 
simulations of \scite{navar95a} (DNF),

\begin{equation}
\rho_{\mbox{\tiny DM}}(r) = \rho_{\mbox{\tiny DM,0}} \left[ x (1+x)^2 
 \right]^{-1}
\end{equation}
where $x = r/r_{s}$ and $r_{s}$ is the scale radius (arcmin); and a Hernquist 
profile (DHF),

\begin{equation}
\rho_{\mbox{\tiny DM}}(r) = \rho_{\mbox{\tiny DM,0}} / \left[ 2 \pi b 
 (1 + b)^{3} \right]
\end{equation}
where $b = r/r_{s}$ and $r_{s}$ is the scale radius (arcmin). 

\subsubsection{Fitting the models}

Determination of the best-fit parameters for a cluster model proceeds in
the way commonly employed for spectral fitting. The fit statistic
employed is maximum likelihood, rather than chi-squared, since the data
are generally strongly Poissonian. The fit and its local slope are
determined at some initial position in the parameter space. This
information is used to predict an improved set of model parameters and
the fit statistic re-determined. The process is iteratively repeated
until the statistic slope falls below a pre-determined value. One
limitation of this method is, however, that the fitting tends to follow
the local gradient in the statistic, until it encounters a minimum. Thus
the fit can become trapped in a ``valley'', which it regards as the
best-fit result, even though a more suitable combination of parameters
may occur elsewhere. To avoid this, we randomly perturb models during
analysis and force them to re-fit (to check if the same minimum is
produced).

Confidence regions can be derived for each best-fit parameter, by
offsetting the parameter of interest from its best-fit value (both above
and below the best-fit), and reoptimising the other parameters. The
resulting increase in the fit statistic, from its optimum value, is used
to determine what offset would need to be applied in order to create a
user-defined change in the statistic. This defines the required
confidence interval. We use the form of the maximum likelihood statistic
introduced by \scite{cash79a}, such that changes in the statistic have
the same significance as changes in chi-squared. Hence, for each
parameter, an increase in the Cash statistic of 1 corresponds to a 68\%
confidence interval, and an increase of 2.71 to 90\% confidence. The
above process is repeated for each model parameter for which an error
estimate is required.

Errors in physical quantities which are functions of radius (such as mass
or temperature) are generally affected by several model parameters. We
derive error envelopes for such quantities by taking the outer envelope
of all of the curves generated by perturbing each free parameter to its
upper and lower error bounds. Because these envelopes are derived using
every parameter combination, each offset to their error bounds, the
result is a {\it conservative} estimate of the statistical uncertainty.

Once the total gravitating mass distribution has been determined (using
Equation~\ref{eq:xray_hydro}, if the mass has not been modelled directly)
the various mass components in the cluster can be separated.  The gas
mass profile is calculated from the fitted parameters. The galaxy mass
profile can be constructed from the observed luminosity profile for the
cluster, assuming a constant mass-to-light ratio. Subtraction of these
components from the total mass profile then yields the dark matter
profile.

\subsection{ROSAT PSPC reduction}

The aim of the ROSAT analysis is to obtain well constrained gas density 
parameters, which can then be utilised in the ASCA analysis.

The raw data, obtained on May 25th 1991, were reduced using the Starlink 
ASTERIX X-ray analysis package. Periods of high background were removed from 
the data, reducing the effective exposure time to 42 ksec but making the 
background subtraction substantially more reliable.

Subtraction of the X-ray background was accomplished by selecting data
from an annulus ($27\arcmin-33\arcmin$), ignoring pixels covered by the
detector support structure or containing point source emission. This
background sample was then extrapolated to cover the whole field, using
the PSPC energy-dependent vignetting function. Since the X-ray surface
brightness profile for A2218 can be traced to a maximum radius of
$12\arcmin$, the chosen background annulus is free of source emission. In
the following analysis, the data are restricted to lie within $9\arcmin$
to avoid possible systematic effects from uncertainties in the background
subtraction at large radius.

\begin{figure}
 \vspace*{8.5cm}
 \includegraphics{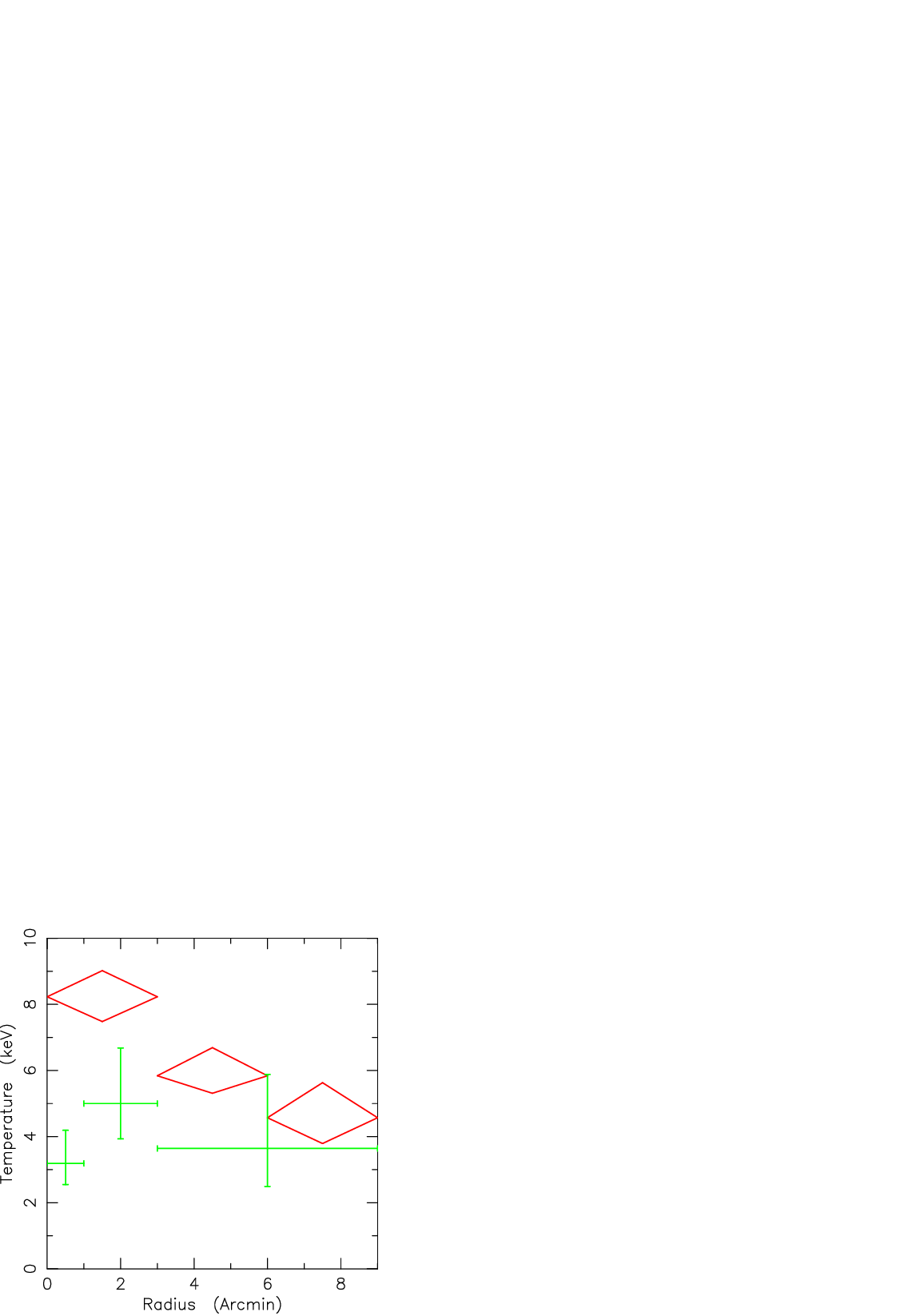}
 \caption[PSPC gas temperatures in projected annuli]{\small Gas temperatures
 derived in projected annuli. The barred crosses are taken from ROSAT PSPC 
 analysis and the diamonds from ASCA GIS analysis. The results suggest 
 a temperature decline with radius, with weak evidence (from the PSPC) for
 central cooling.}
 \label{fig:temp_2d}
\end{figure}

The exposure-corrected, background-subtracted PSPC data were summed to
provide an integrated cluster spectrum and split into concentric annuli,
centred on the cluster core, within which spectra were extracted. Fitting
these spectra gives an indication of the depth of the cluster potential
well and the radial structure of the cluster gas density and temperature
parameters (see Fig~\ref{fig:temp_2d}). However, any gradients present in
these distributions will tend to be underestimated, due to the smoothing
effects of the instrument psf and projection along the line-of-sight.
Using annuli of width greater than the instrument psf minimizes the
former effect.

In order to create a spectral image dataset, allowing a full spatial and 
spectral analysis, the data were formed into images of channel width 10, over 
the channel range 11-230 (approximately 0.2-2.3keV). This results in a data 
cube, from which specific regions can be selected and analysed.

Within the cluster emission there is one bright source contaminating the
data, located at a radius of $11.1\arcmin$ from the cluster centre. In
the PSPC analysis, the point source can be eliminated by ignoring the
data collected in that region. However, this is not possible with ASCA,
since the extended psf ensures that it is not discernable as a discrete
source. To determine whether the source significantly affects our
analysis, we model it using the PSPC data. The cluster emission is first
fitted with the point source pixels removed. This model of the cluster
emission is then subtracted from the original data, leaving behind just
the point source, which is fitted with a power-law spectral model.

The fitted index, 1.25, indicates the softness of the source - the
majority of the flux is emitted below 0.5keV. This is consistent with
identification of the source as SAO17151, a bright star with a soft
spectrum \cite{markev97a}. If the PSPC-determined model for the point
source is subtracted from the ASCA data, the cluster fits are modified to
the extent that a 1.5\% difference in the total gravitating mass at 2Mpc
results. This effect is negligible compared to other errors, so no
attempt has been made to remove the source from the ASCA data.

\subsection{ASCA GIS reduction}

The ASCA analysis aims to constrain the gas temperature and metallicity
profiles, using the PSPC derived gas density profile. A2218 was observed
by the ASCA X-ray telescope on April 30th 1993. In this paper we use only
data from the two gas imaging spectrometers (GIS2 and GIS3), since these
have a wider field of view ($\sim 50\arcmin$ diameter) and greater
high-energy detector efficiency (up to 10keV) than the CCD detectors
(SIS0 and SIS1). An additional reason is that an accurate model for the
large, asymmetrical and energy dependent psf is available for the GIS
detectors, constructed from Cyg X-1 observations. These restrict analysis
to a maximum radius of 18$\arcmin$ and an energy-range of 1.5-11 keV
\cite{takah94a}, which is not a significant limitation in the case of
A2218.

Standard procedures for ASCA analysis, followed in this paper, are given
by \scite{day95a}. The recommended screening criteria are applied to the
raw data, removing data taken during times of high background flux.
Subtraction of the X-ray background is complicated by the telescope psf.
This has the effect of ensuring that no region of the detector is free
from source flux. Hence, we extract an ``average'' background dataset
from the publicly distributed set of blank-sky pointings \cite{day95a}.
These datasets suffer from mild point-source contamination, as sources
are not completely averaged out, but are currently the best available
solution.

The results of a naive annular spectral analysis of the ASCA data are
shown in Fig~\ref{fig:temp_2d}. However, it is important to bear in mind
the limitations of this approach. Cross-talk between annuli is
significant \cite{takah94a}, and the energy-dependent spreading of flux
results in a distorted temperature profile, such that analysis of a
simulated isothermal cluster would give a temperature which rises with
radius. In practice, the temperature appears to {\it decline} with
radius, indicating that a real gradient is present.

For 3D analysis, spectral-image datasets with contiguous energy bands of
width 50 raw channels are created. Since the cluster centre is offset
from the detector centre, data beyond a radius of 9$\arcmin$ are not
fitted (the offset added to the radial extent of the source is similar to
the maximum radius where psf calibrations apply). This restriction
minimises the effects of poor calibration and high background near the
detector edge. As both GIS instruments behave similarly, the datasets are
fitted simultaneously. The Cash statistic has been used to identify
best-fit models, but similar results for best-fit parameters and for
comparison between the quality of fit of different models, is obtained
using the $\chi^{2}$ statistic.

\section{Results}
\label{sec:2218_3dres}

We first compare the results of our analysis with published studies, to
investigate whether the fitted models are consistent with earlier work on
A2218.

Integrated spectral analyses of clusters produce ``mean'' quantities
which are representative of the entire object. Assuming isothermality,
\scite{mchard90a} derived a gas temperature of $6.72^{+0.5}_{-0.4}$ keV
together with an iron abundance of $0.2^{+0.2}_{-0.2}$ Z$_{\odot}$ (using
an RS emission code, where all other heavy element abundances were fixed
at 0.5 Z$_{\odot}$). This is in agreement with an earlier, much less well
constrained examination \cite{perr81a}. 
More recently, \scite{mushot97a} have derived $T=7.2$~keV and
$Z=0.18\pm0.07$ Z$_{\odot}$, from an integrated ASCA spectrum.

Fitting an RS model to our ROSAT data results in a gas temperature of
$4.7^{+1.1}_{-0.9}$ keV and a hydrogen absorption column of
$2.6^{+0.2}_{-0.1}$x$10^{20}$ cm$^{-2}$ (with metallicity fixed at the
Ginga value). This absorption agrees with the Stark level of
$2.58^{+0.18}_{-0.18}$x$10^{20}$ cm$^{-2}$ \cite{stark92a}. The
temperature is lower than the Ginga result of \scite{mchard90a}, but the
energy range of the PSPC is not very suitable for determining the
temperature of such hot gas. It has been found previously
\cite{markev97b} that PSPC results tend to be biased low for high
temperature clusters. Fitting an integrated spectrum simultaneously to
the GIS2 and GIS3 data gives a gas temperature of
$6.73^{+0.46}_{-0.44}$keV and a metallicity of
$0.20^{+0.08}_{-0.08}$Z$_{\odot}$ (with the hydrogen column fixed at the
PSPC value), in good agreement with the Ginga and \scite{mushot97a} 
results.

The ability to extract and analyse spectra from independent regions of
the cluster represents a significant advance over analysis of the
integrated emission. Fig~\ref{fig:temp_2d} shows the derived annular
temperature profiles from both ROSAT and simultaneous GIS2/GIS3 analysis.
The PSPC results are consistent with \scite{sidd95a}, with a possible
temperature drop visible in the central bin. However, the evidence for
central cooling is statistically rather weak, and the derived central
cooling time of $\sim1.5$x$10^{10}$yr is comparable with the Hubble time,
so a strong steady-state cooling flow appears to be ruled out.

If the hydrogen column is fitted, using the PSPC data, it is found to be
consistent with the Stark value \cite{stark92a} throughout the cluster,
apart from a slight rise in the centre. This may be due to matter
deposited by an earlier, disrupted cooling flow (as noted by
\pcite{sidd95a}). The ASCA analysis suggests that the metallicity may be
slightly lower than the integrated value in the cluster centre with a
shallow radial rise. However, all points are consistent with the
\scite{mchard90a} value of 0.2 Z$_{\odot}$.

Analysis using spectral-image datasets allows extraction of the 3D gas
density and temperature distributions within A2218. In the subsequent
cluster analysis, both `temperature models' (fitting for $\rho_{gas}(r)$
and T$_{gas}(r)$) and `mass models' (fitting for $\rho_{gas}(r)$ and
M$_{tot}(r)$) are used. Parameters representing the metallicity and
cluster position are also fitted.

The best constrained parameters derived from PSPC analysis pertain to the
shape of the gas density distribution. Comparable analyses have been
carried out using Einstein (\pcite{perr81a}; \pcite{boyn82a};
\pcite{birk94a}), and ROSAT \cite{sidd95a} data. These studies agree
that when a King profile is assumed, A2218 is well modeled with a core
radius, $r_{c}$, of $\sim 1\arcmin$ and an index, $\alpha_{\rho}$, of
$\sim 1$ (equivalent to a $\beta$-value of 0.67). Higher resolution
analysis, using the ROSAT HRI, has been performed by \scite{squire96a}
and \scite{markev97a}. These studies detect the presence of smaller scale
structure, with three surface brightness peaks visible within the central
arcminute, none of which coincides with the central cD galaxy. This
structure cannot be resolved with either the PSPC or GIS detectors, and
indicates that the core of A2218 departs from a fully relaxed state.

Since ASCA is poor at constraining the gas density distribution for such
a compact source, the PSPC fit values for core radius and index are
carried over into the ASCA spectral-image analysis. To obtain the
appropriate density parameters, a linear temperature ramp model is fitted
to the entire PSPC data within a radius of 9 arcmin, with the temperature
parameters fixed at those derived from an initial fit to the GIS data.
This model is then re-fitted to the GIS data, complete to a radius of 9
arcmin, with the gas density core radius and index fixed, allowing a fit
of the temperature parameters. In an iterative process, this model is
alternately fitted to the PSPC and GIS datasets until no further change
is observed in the parameter values. With consistency achieved, the
``standard'' gas distribution for A2218 is determined to be $r_{c}$ =
$0.91^{+0.03}_{-0.03}$ arcmin, $\alpha_{\rho}$ = $0.96^{+0.02}_{-0.01}$,
in good agreement with the comparable results discussed above. If an NFW
profile is assumed, the required scale radius, $r_{s}$, is
$10.27^{+0.21}_{-0.20}$ arcmin. However, the NFW parameterisation is
neither preferred nor disallowed by the PSPC data, so we only use a King
parameterisation for the gas density distribution in the following
analysis.

\subsection{All cluster data}

\begin{figure*}
 \vspace*{9.5cm}
 \includegraphics{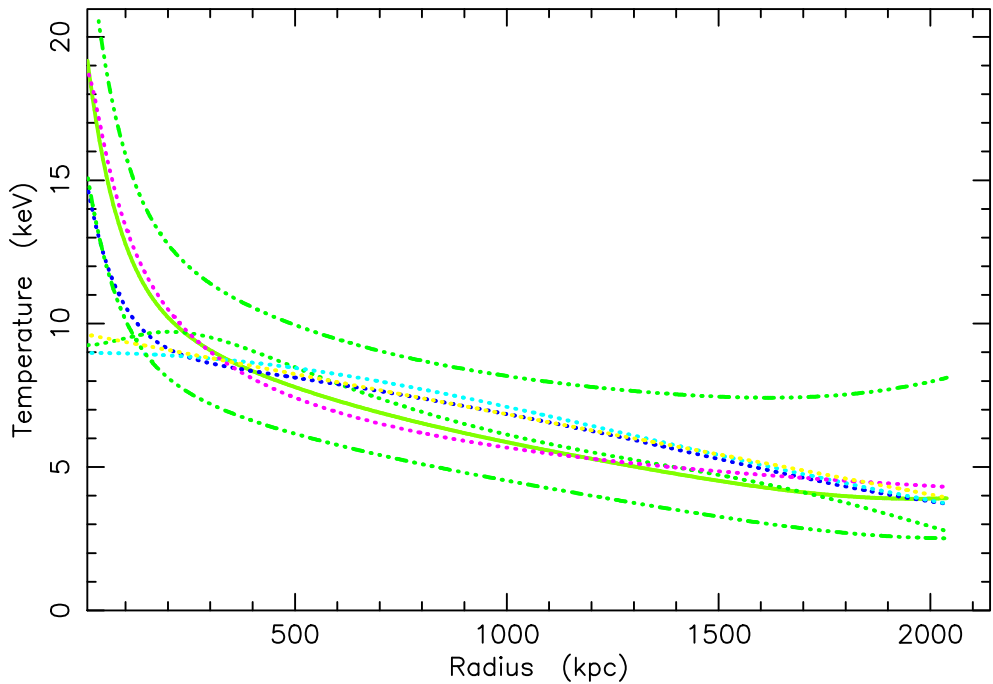}
 \caption[Fitted gas temperature profiles]{\small Fitted 3D temperature 
 profiles for the cluster gas. The MMM (solid) is plotted together with its 
 90\% error envelope (dash-3dot) and the remaining best-fit models (dotted). 
 All of the profiles feature gas temperatures which decrease by a factor of 
 $\sim2$ or more within the region of analysis, and are consistent beyond the 
 cluster core.}
 \label{fig:temp_fits}
\end{figure*}

Fixing the gas density shape parameters at these King values, a range of
models were fitted to the ASCA spectral-image data. On the basis of their
Cash statistic, a set of models best-fitting the observed cluster data
within a radius of 9 arcmin is selected (Table~\ref{table:best_mod} lists
the main parameters for each model) as representative of A2218. As can be
seen from the Table, the isothermal model is a significantly poorer fit
to the data. It is included for comparison with the other models. All of
the temperature profiles obtained for this set (apart from that for the
isothermal model) are plotted in Fig~\ref{fig:temp_fits}. Beyond the
central arcminute ($\sim$230 kpc) the profiles are in good agreement and
quite non-isothermal. The typical 90\% confidence envelope for an
individual model (remember that these are conservative envelopes)
generally encompasses the spread of these best-fit profiles (a single
error envelope is included to illustrate this point). Within the central
region, which lies within a single ASCA psf, a greater spread in
temperature is allowed by the data. However, despite this divergence in
temperature at small radius, all of the best-fit models, bar the
isothermal model, have similar Cash statistics (see
Table~\ref{table:best_mod}).

The abundance of heavy elements has been assumed, in all of the above
models, to be constant over the cluster. However, the spectral
capabilities of ASCA allow us to test this assumption. Allowing a linear
metallicity gradient with the best-fit linear temperature ramp model
gives a slope of $2.8_{-6.5}^{+4.5}$ x $10^{-2}$ Z$_{\odot}$
arcmin$^{-1}$, a value consistent with uniform metallicity. The effect of
this best-fit slope upon other model parameters is negligible, hence
freezing the metallicity gradient at zero does not bias our analysis.

\subsection{Central cluster data}

We now examine the claim that lensing analyses require a larger cluster
mass than is consistent with the X-ray data. The model which provides the
greatest gravitating mass at the critical radius ($\sim$85 kpc) is
therefore selected for examination. This model, the DHF best-fit, using a
Hernquist description for the DM distribution and a King description for
the gas density profile (see Table~\ref{table:best_mod}), implies a high
central gas temperature. In the following comparisons, we refer to this
model as the ``maximum-mass model'' (MMM).

The MMM temperature profile features a factor of 2 rise within the
central 1$\arcmin$. This raises two questions: does such a steep
temperature gradient raise physical problems (would it be convectively
unstable?), and is it consistent with the X-ray spectral data observed
within the central region? We will return to the first question in
Section~\ref{sec:2218_entros}.

\begin{figure}
 \vspace*{8.5cm}
 \includegraphics{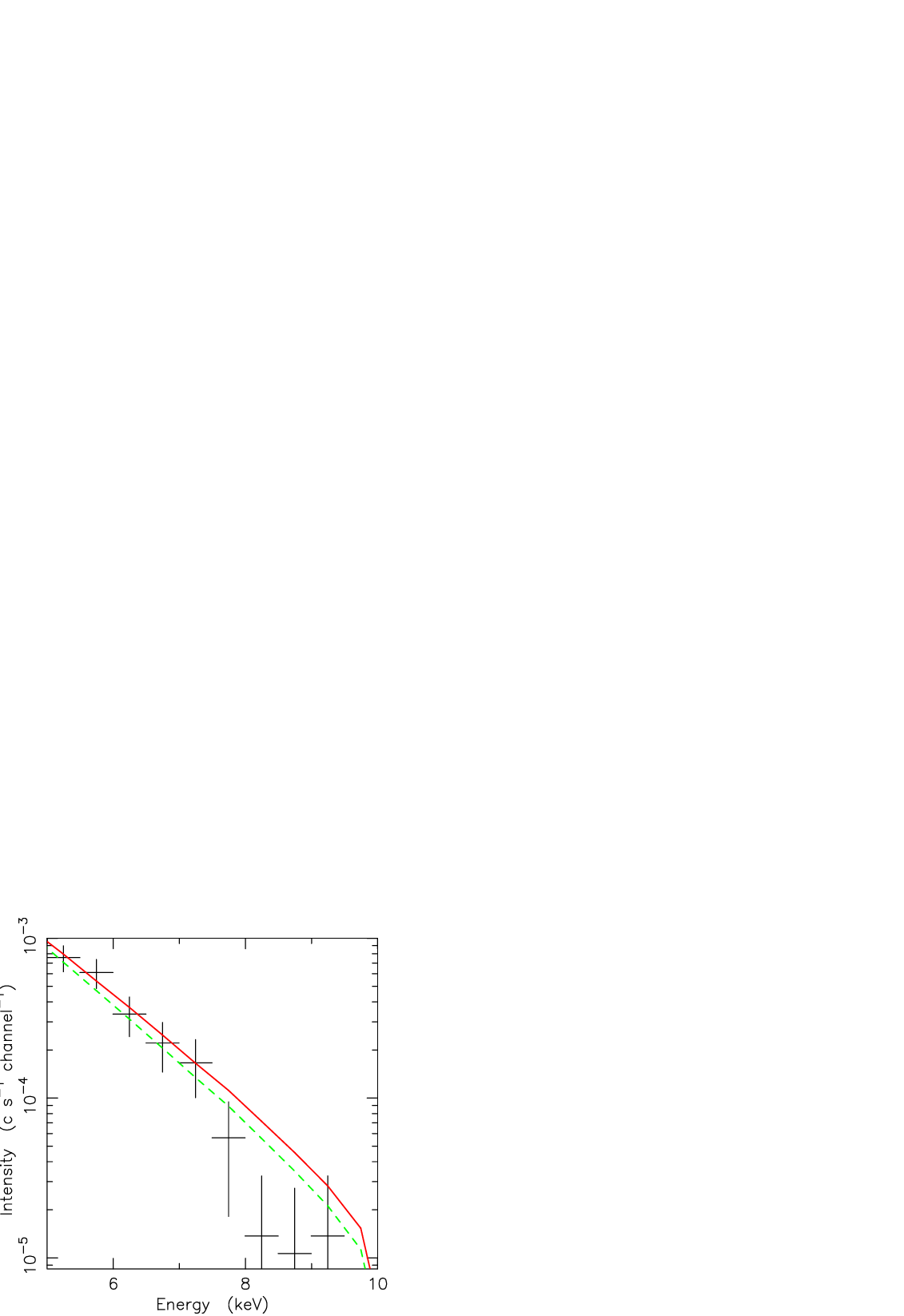}
 \caption[Hard-tail spectral test (3D)]{\small The GIS3 spectrum from the 
central 1$\arcmin$ (crosses) is compared to the prediction from the MMM model
(solid) and the isothermal model (dashed). The MMM model provides more flux 
at high energy, because of the centrally increasing temperature, but the 
difference is very small. Both models are consistent with the observed data.}
 \label{fig:g3_ispec}
\end{figure}

To address the latter question, GIS3 spectra integrated within
$r=1\arcmin$, where the MMM temperature rises steeply, are compared with
the predictions of the MMM and isothermal models in
Fig~\ref{fig:g3_ispec}. For clarity, only the energy range of 5-10 keV is
displayed, since this is where the impact of very hot gas will be most
apparent. Although the MMM provides a reasonable fit to the spectral
imaging data {\it as a whole}, it does not follow that the data in the
central regions need be consistent with the high model temperature. In
practice, for both instruments (while the GIS2 spectra are not shown in
Fig~\ref{fig:g3_ispec}, they behave similarly to the GIS3 spectra) the
MMM is a good match to the data. The isothermal model is, however, also
consistent with the restricted dataset. The conclusion from this is that
while the data do not rule out a central temperature rise,
they do not require one either. The reason for this is that within the
ASCA bandpass, plasmas with temperatures of 8 and 18keV do not have
substantially different spectral signatures and are thus difficult to
differentiate (this is analogous to the difficulty that the PSPC has in
dealing with clusters hotter than 2-3keV). This problem is compounded by
the limited spatial resolution of ASCA.

\begin{table}
\centering
\begin{tabular}{|c|c|c|c|}
\hline
\multicolumn{1}{|c|}{Model} &
 \multicolumn{1}{|c|}{Central} &
  \multicolumn{1}{|c|}{T$_{gas}$} &
   \multicolumn{1}{|c|}{Relative} \\ 
Form       & T$_{gas}$ & Gradient  & Cash    \\ 
           & (keV)     & (various) & Statistic \\ \hline
LTF        & 9.64      & 0.63      & -636.08 \\
LTF (ISO)  & 7.90      & zero      & -626.85 \\
TTF        & 19.01     & 1.20      & -634.53 \\
KTF        & 8.98      & 3.18      & -635.96 \\
DNF        & 15.59     & n/a       & -636.12 \\
DMF        & 9.24      & n/a       & -635.20 \\
DHF (MMM)  & 20.42     & n/a       & -635.33 \\ \hline
MMMC       & 16.44     & n/a       & -632.50 \\
\hline
\end{tabular}
\caption[Best-fit model parameters]{\small Temperature and 
 statistic parameters 
 for a variety of fitted 3D models are shown. Note that the MMMC is not a 
 best-fit model in the same sense as the others, as is described in the text. 
 For all models, the gas density core radius, r$_{c}$, and index, 
 $\alpha_{\rho}$, are fixed at the values determined from the ROSAT data.
 Explanations for the model acronyms are given in the text.}
\label{table:best_mod}
\end{table}

\section{Comparison of X-ray and lensing masses}

Deep optical observations of A2218 reveal a number of major arcs and a
wealth of minor arclets. These features are the result of gravitational
lensing, an effect which is independent of the physical state of the
cluster gas. Instead, uncertainty lies in the characteristics of the
background galaxies and the possibility of matter sub-clumps along the
line-of-sight.

Using the models fitted to A2218, we can derive projected gravitating
mass profiles, suitable for comparison with the results of lensing
analysis. This involves assuming a maximum outer radius for the cluster
mass distribution. In the following analysis we take this to be the
maximum radius of the data used in cluster fitting, which is $9\arcmin$
($\sim$2 Mpc). The choice of projection radius has a minor impact upon
the derived projected mass profiles (compared to other uncertainties) so
long as the chosen radius is sufficiently large, $\geq2$Mpc. For example,
the difference in projected mass within 2Mpc, between models with maximum
radii of 2.0 and 3.0 Mpc, is 9\%.

Since lensing analysis measures the matter distribution on both small
(strong lensing occurs where the surface mass density is high) and large
(weak lensing is theoretically observable to the edge of the cluster)
scales, comparison with X-ray results is extremely informative. In
Fig~\ref{fig:mass_predout} the X-ray derived projected mass profiles for
the MMM and isothermal model are plotted, together with the lensing
results of \scite{kneib95a}, \scite{loeb94a} and
\scite{squire96a}.

\subsection{Strong lensing}

Strong lensing analysis has been performed, using spectroscopically
observed arcs, by a variety of groups (\pcite{loeb94a}; \pcite{kneib95a};
\pcite{kneib96a}; \pcite{saran96a}; \pcite{natar96a}). These analyses
provide the gravitating mass within the critical radius associated with
the formation of giant arcs, which in A2218 is $22.1\arcsec$ ($\sim
85$kpc).

\begin{figure*}
 \vspace*{9.5cm}
 \includegraphics{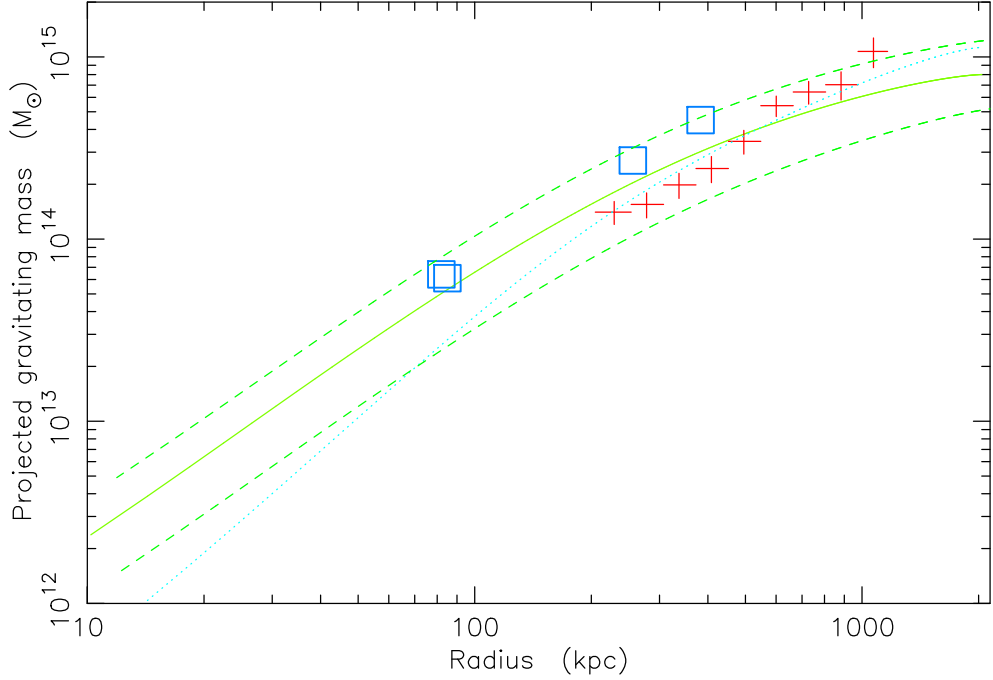}
 \caption[Projected gravitating masses]{\small Projected gravitating mass 
 profiles for the MMM (solid line) and isothermal (dotted line) are plotted.
 Also shown is the 90\% error envelope for the MMM (dashed lines). 
 Overlaid are 
 strong lensing points from \scite{kneib95a} and \scite{loeb94a} (boxes) 
 together with the weak lensing points of \scite{squire96a} (crosses).}
 \label{fig:mass_predout}
\end{figure*}

The two strong lensing points, plotted in Fig~\ref{fig:mass_predout} at
this radius, are consistent with a projected mass of
$\sim$6x10$^{13}$M$_{\odot}$, despite their use of different mass
distributions (\pcite{kneib95a} assume a bipolar mass model while
\pcite{loeb94a} use an isothermal sphere distribution). The corresponding
projected mass for the MMM is 5.3$^{+1.2}_{-4.1}$x10$^{13}$ M$_{\odot}$.
The uncertainty associated with this mass is large enough to include the
lensing derived value, so, within the calculated errors, the X-ray and
strong lensing analyses are consistent, when the MMM is used. The
corresponding mass from the isothermal model is
2.8$^{+0.2}_{-0.2}$x10$^{13}$ M$_{\odot}$ (see
Fig~\ref{fig:mass_predout}), in agreement with the factor of 2
discrepancy reported by previous analyses (\pcite{loeb94a};
\pcite{miral95a}; \pcite{kneib95a}).

The bipolar mass model of \scite{kneib95a} was used by these authors to
extract a further two masses (see Fig~\ref{fig:mass_predout}), at the
radius of the second mass clump (256 kpc) and at the maximum radius where
giant arc constraints can be applied (383 kpc). The projected, enclosed
gravitating mass at these radii for both the MMM and the isothermal model
are given in Table~\ref{table:strong_lens}. No error estimates are given
by \scite{kneib95a} for the lensing masses. However, it can be seen that
both of these outer lensing points lie just outside the statistical error
envelope of the MMM, unlike the mass derived at the critical radius. This
is due to the MMM profile flattening at large radius. Note however that
both of the outer predicted masses depend upon the bimodal model of
\scite{kneib95a}, whereas the inner point can be derived in a
model-independent fashion.

\begin{table*}
\centering
\begin{tabular}{cccc}
\hline
Source/Mass    & 85kpc  (x10$^{13}$M$_{\odot}$) 
 & 256kpc (x10$^{14}$M$_{\odot}$) & 383kpc (x10$^{14}$M$_{\odot}$)\\ \hline
Kneib 1995     & 6.1 & 2.7 & 4.5 \\
Loeb\&Mao 1994 & 6.4 & -   & -   \\
MMM            & 5.3$^{+1.2}_{-4.1}$ & 2.0$^{+0.5}_{-1.6}$ & 
 3.0$^{+0.7}_{-2.4}$ \\
Isothermal     & 2.8$^{+0.2}_{-0.2}$ & 1.6$^{+0.2}_{-0.1}$ & 
 2.8$^{+0.1}_{-0.1}$ \\
\hline
\end{tabular}
\caption[Projected gravitating masses]{\small Projected gravitating 
 masses from 
 strong lensing and X-ray analysis are compared at the radii used by
 \scite{kneib95a}. At the critical radius the isothermal mass is a factor of 
 2 too low, although the discrepancy reduces greatly at larger radii.}
\label{table:strong_lens}
\end{table*}

The \scite{kneib95a} bimodal mass model has received additional support
from detailed optical observations by \scite{ebbel96a}. In this analysis,
the redshift of one of the faint arcs was determined for the first time
and found to be in good agreement with the value predicted by
\scite{kneib95a}.

The conclusion which can be drawn from this X-ray/strong lensing
comparison is that it is {\it not} clear that the results from the two
approaches are inconsistent. We have shown that when a model such as the
MMM is used, which includes a central mass cusp, the predicted masses at
the critical radius agree within the X-ray statistical error envelope
(see Table~\ref{table:strong_lens}). If the gas is assumed to be
isothermal, the previously noted discrepancies can be reproduced (see
Fig~\ref{fig:mass_predout}).

\subsection{Weak lensing}
\label{sec:2218_wlens}

A statistical analysis of the weakly lensed arclets has been carried out
by \scite{squire96a}, allowing the slope of the surface mass density to
be mapped within a radius of $\sim 1$Mpc. Fig~\ref{fig:mass_predout}
shows that, within errors, the magnitudes of the weak lensing points are
consistent with the MMM. However, the trend indicated by these points is
for a steeper profile, providing more gravitating mass at large radii
than predicted by the MMM.

A significant difficulty with the weak lensing analysis, noted by
\scite{squire96a}, is the procedure used for normalisation. This is done by
defining a reference annulus at large radius within which the cluster mass 
contribution is assumed to be negligible. Since the annulus used by 
\scite{squire96a} has an inner radius of 800 kpc, while the X-ray surface
brightness profile can be traced to $r>2$~Mpc, some shear signal from the
cluster must actually be present within the reference annulus. Hence the
recovered normalisation is an underestimate, such that the weak lensing
mass profile can only be regarded as a lower bound to the projected mass.
The impact of this is examined further in
Section~\ref{sec:2218_cegal}.The correction for this effect has been
estimated by \scite{squire96a} to be a factor of $\sim$1.2-1.6 in
projected mass.

If the weak lensing points are adjusted to take account of this factor,
this has two important implications for the MMM comparison. First, the
normalisation of all data points increases, bringing the inner points
into better agreement with the MMM profile while the outer points move
further from consistency. Second, because this adjustment is a DC effect
for the reconstructed projected surface mass density (of the cluster), it
does not act equally on the radially integrated points. Hence, when the
normalisation is raised, the slope of the weak lensing mass profile in
Fig~\ref{fig:mass_predout} {\it increases}, bringing it into greater
conflict with the MMM profile. Thus the flatter slope of the MMM cannot
be made consistent with the weak lensing points by shifting the reference
annulus to larger radii.

Even when the normalisation adjustment is applied, this is insufficient
to ensure full consistency with the outer strong lensing points of
\scite{kneib95a}, which (see Table~\ref{table:strong_lens}) lie above the
values of 1.5x10$^{14}$ M$_{\odot}$ and 2.5x10$^{14}$ M$_{\odot}$
predicted at the same radii by weak lensing. There is, however, a well
understood effect whereby the weak lensing signal is suppressed at small
radii due to contamination by cluster galaxies, reducing the derived weak
lensing mass (\pcite{kaiser93a}; \pcite{squire96a}). Correcting for this
would bring the inner weak lensing points points into greater consistency
with both the strong lensing and X-ray results.

Overall, then, this comparison indicates that at large radii, $>250$ kpc,
the weak lensing and X-ray analyses are reasonably consistent, though the
weak lensing results tend to give more mass at large radii. Further work
(observations of increased numbers of arclets over a wider field) is
required to reduce the uncertainty in the weak lensing analysis and to
move the reference annulus to larger radii, where the cluster mass
contribution is lower.

\section{Gas entropy}
\label{sec:2218_entros}

\begin{figure*}
 \vspace*{9.5cm}
 \includegraphics{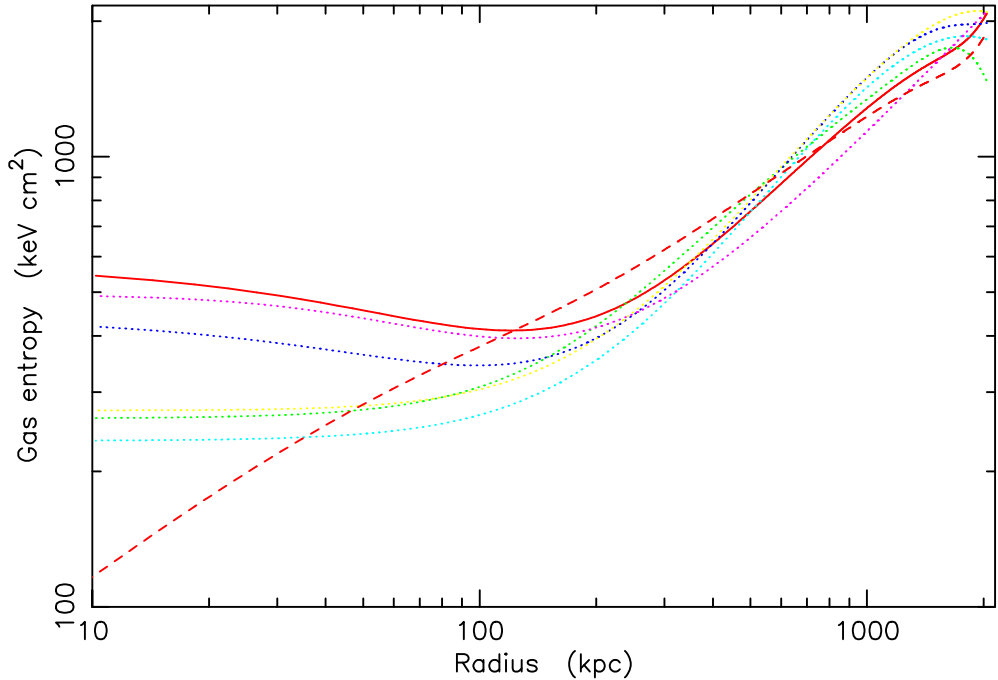}
 \caption[Derived gas entropy profiles]{\small Derived entropy profiles
 for the cluster gas. The MMM is plotted for both a King (solid) and an NFW
 (dashed) gas density distribution. The remaining best-fit models (dotted) 
 are plotted using the King parameterisation. 
 It can be seen that all of the models are consistent with a gas 
which is convectively stable, except within $\sim 120$kpc, where the high 
central gas temperatures of certain models (such as the MMM, see 
Table~\ref{table:best_mod}) lead to an increase in entropy. Where the
NFW gas density form is used, the derived entropy increases at all radii.}
 \label{fig:egas_fits}
\end{figure*}

In Fig~\ref{fig:egas_fits} the derived gas entropy profiles for the
best-fit cluster models are plotted. These indicate that beyond a radius
of $\sim120$kpc the entropy increases with radius, as is expected for gas
which is convectively stable. However, within this radius three of the
best-fit models (including the MMM) exhibit a slight inward rise in
entropy, making the gas convectively unstable. For the MMM, the predicted
increase in entropy which occurs between a radius of 120kpc and the
cluster centre is $\sim30\%$.The reason for this behaviour is that the
temperature increases rapidly at small radii while the gas density is
forced to flatten (due to the use of a King model description). However,
it is precisely this temperature increase that allows the model to
achieve consistency with the gravitational mass derived from strong
lensing.

As noted in Section~\ref{sec:2218_3dres}, the NFW gas parameterisation is
neither preferred nor disallowed by the PSPC data. However, if the King gas 
density parameterisation used for the MMM is replaced by the NFW 
parameterisation, the central entropy drops by $\sim75\%$ (See 
Fig~\ref{fig:egas_fits}). This difference is large enough to ensure that 
entropy rises continuously at all radii, removing the problem of convective 
instability.

\section{Central galaxy mass}
\label{sec:2218_cegal}

An alternative method for bringing the strong lensing and X-ray analyses
into agreement has been examined by \scite{makin96a}. In this study, a
massive central galaxy was embedded within the cluster to increase the
predicted core mass without violating observational X-ray constraints. As
the cD galaxy envelope can be traced to at least $25\arcsec$ (96kpc),
which encompasses the critical radius, this is potentially an important
consideration.

\scite{makin96a} tested this hypothesis by, firstly, constructing a total mass
distribution which included both cluster and cD components. These were
parameterised by a King mass model (equation~\ref{eq:rho_dm} with
$\alpha_{\rm DM}=3/2$) and an isothermal sphere mass model ($\alpha_{\rm
DM}=1$), respectively. Secondly, the gas temperature distribution was
constrained to be isothermal at large radius (beyond the cD galaxy) and
to rise or fall linearly with $r$ at small radius. By assuming the gas to
be in hydrostatic equilibrium, the gas density distribution corresponding
to different temperature models was extracted and compared to the
observed Einstein surface-brightness data.

Constraining the temperature to be isothermal at $r>31\arcsec$, at the
\scite{mchard90a} derived value, \scite{makin96a} found that models
consistent with both the X-ray surface brightness and with enough central
mass to account for the strongly lensed arcs had two notable features.
Firstly, a cD component was required to ensure that the critical surface
mass density was attained by the model. Secondly, the gas temperature
rose sharply within $31\arcsec$, reaching a central value of $\sim11$keV.
Models which did not include a cD mass contribution, or which assumed the
gas to be isothermal at all radii, were unable to provide sufficient mass
at small radii to account for the lensing data, whilst remaining
consistent with the Einstein data.

The projected gravitating mass within 1.5 Mpc, from the model favoured by
\scite{makin96a}, of 1.3x10$^{15}$ M$_{\odot}$ is in reasonable agreement
with the MMM value of 9.0$^{+1.4}_{-5.3}$x10$^{14}$ M$_{\odot}$. This
consistency is also found at smaller radii, r=100 kpc, where
\scite{makin96a} obtains a mass of 5.2x10$^{13}$ M$_{\odot}$ compared to
6.6$^{+1.5}_{-5.1}$x10$^{13}$ M$_{\odot}$ for the MMM. It should be noted
that the MMM employs a mass profile with a central mass cusp
($\rho_{\mbox{\tiny DM}} \propto r^{-1}$) but does not include any
discrete component corresponding to the central galaxy. We now explore
the effect of adding such an additional central component to our model.

Adding a central galaxy of radius 100kpc and adjustable mass
normalisation to the MMM, we find that the statistical quality of the fit
deteriorates. The largest additional mass allowed, at the 95\% level, is
1.7 x 10$^{12}$ M$_{\odot}$ (within 100kpc). From here on this model is
referred to as the MMMC, since it represents the MMM plus a central
galaxy mass. Note however that the MMM, which contains no central mass
component, is still statistically preferred (see
Table~\ref{table:best_mod}).

This central mass is substantially less than the \scite{makin96a} value
of 5.2x10$^{13}$M$_{\odot}$ (recall, however, that our cluster profile
contains a central cusp, whilst Makino's has a flat core). Compared to
the MMM, it provides a mass increase at the critical radius of only
$\sim$5\%. However, whilst the addition of a central mass component has
little impact at small radius, it has the rather counter-intuitive effect
of providing more mass at {\it large} radius. The reason for this is
that, in the case of the MMM, the DM distribution has a scale radius of
1.89$^{+0.06}_{-1.16}$ arcmin. When a central mass component is added,
the DM profile no longer needs to peak so sharply and its scale radius
increases to 6.79$^{+0.95}_{-1.49}$ arcmin.

\begin{figure*}
 \vspace*{9.5cm}
 \includegraphics{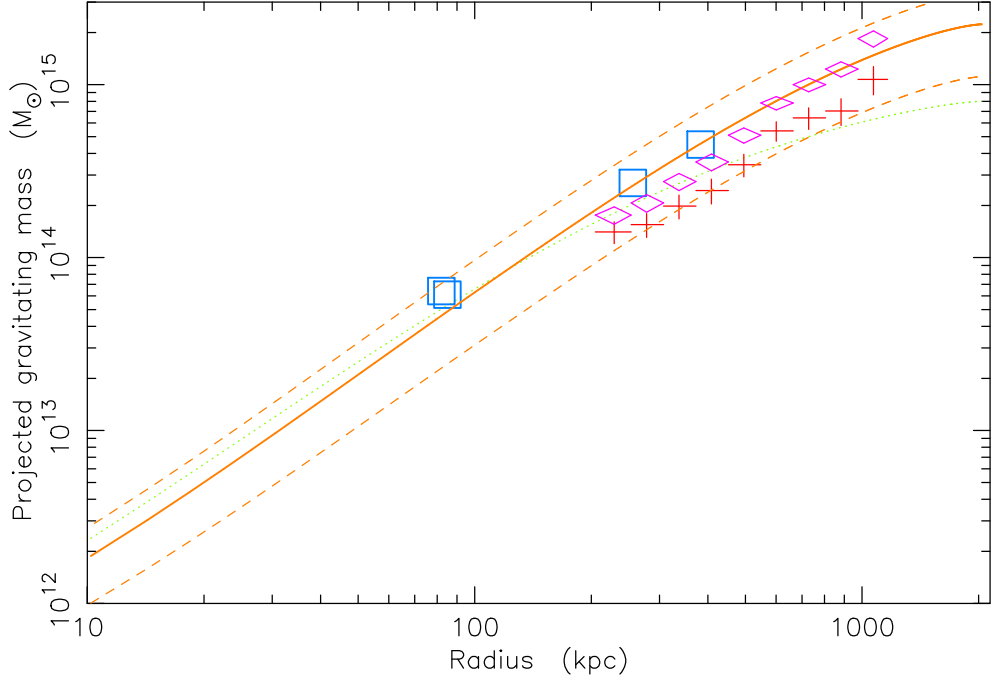}
 \caption[Projected gravitating masses]{\small The projected gravitating 
 mass profile for the MMM (dotted line) is compared with that of the MMMC 
 (solid line). Also shown is the 90\% error envelope for the MMMC (dashed 
 lines). Overlaid are the strong lensing points from \scite{kneib95a} and
 \scite{loeb94a} (boxes) together with the weak lensing points of 
 \scite{squire96a} (crosses). Also shown are the weak lensing points after
 correction for cluster mass (as predicted by the MMMC) within the lensing
 reference annulus (diamonds). It can be seen that the MMMC provides a good 
 match to both the strong and weak lensing points (especially after 
 re-normalisation of the latter).}
 \label{fig:mass_mmmc}
\end{figure*}

Fig~\ref{fig:mass_mmmc} shows the MMM and MMMC mass profiles compared to
the strong and weak lensing results. It can be seen that the increased
mass of the MMMC at large radius (compared to the MMM) allows it to be
consistent, within its 90\% statistical confidence envelope, with both
the two outer points derived by \scite{kneib95a} and the weak lensing
points of \scite{squire96a}.

We can use our models to estimate the baseline error involved in the weak
lensing analysis, as a result of cluster mass residing within the
reference annulus employed by \scite{squire96a}. In
Fig~\ref{fig:mass_mmmc} the result of using the MMMC to correct the weak
lensing results for this mass is also plotted. It can be seen that the
points come into better agreement with the MMMC profile, particularly at
r$\gtapp500$ kpc. At smaller radii the weak lensing points still lie
systematically below the MMMC profile, although this is to be expected
(see Section~\ref{sec:2218_wlens}) due to dilution of the lensing signal
by cluster galaxies.

\begin{figure*}
 \vspace*{9.5cm}
 \includegraphics{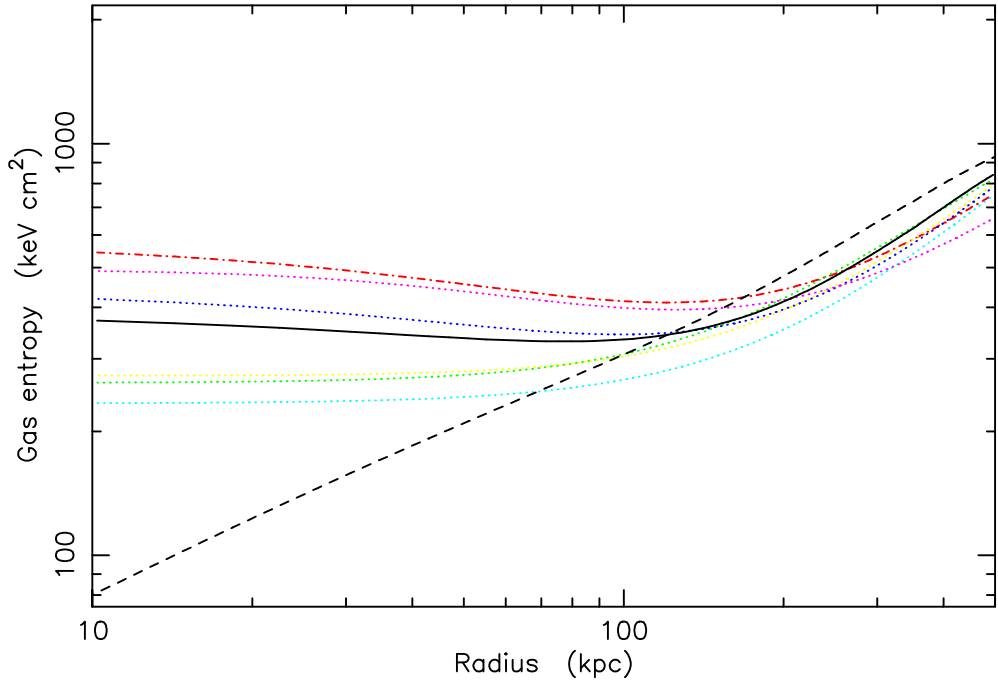}
 \caption[Derived gas entropy profiles]{\small Derived gas entropy profiles
for the inner 500 kpc region are plotted. The MMMC (solid line) can be
seen to provide a considerably flatter central entropy profile than the
MMM model (dash-dot line), where both use a King model for the gas
density. If an NFW gas density form is used with the MMMC (dashed line),
the derived entropy can be seen to increase at all radii. The remaining
best-fit model profiles (dotted) are included for comparison.}
 \label{fig:egas2_fits}
\end{figure*}

In Fig~\ref{fig:egas2_fits} the derived gas entropy profile for the MMMC
is compared with those from the best-fit X-ray models within the central
500kpc region, where the MMM shows a noticeable rise. It can be seen that
the MMMC provides a flatter central entropy distribution, such that the
gas is less likely to be subject to convective instability.

\section{Analysis of galaxy motions}

The dynamics of cluster galaxies provide a further way of investigating
the mass distribution in clusters. In the case of A2218, galaxy redshifts
are available only within a few core radii of the centre. However this
includes the region where the lensing analysis of \scite{kneib95a}, and
the high resolution X-ray observations of \scite{markev97a}, suggest that
the potential may be seriously disturbed.

Galaxy velocity and position data have been obtained from the NASA
Extragalactic Database. This consists largely of the data obtained by
\scite{lebor92a}, who performed an extensive photometric survey of the
cluster core. On the basis of the $3\sigma$ clipping technique of
\scite{yahil77a}, 49 of the 53 galaxies with measured redshifts are
identified as cluster members. All of these galaxies lie within
$3\arcmin$ of the X-ray centroid of the cluster, which has been adopted
as the centre for this optical analysis. The average line of sight
velocity dispersion, calculated using the 49 cluster galaxies, 
and corrected \cite{harrison74a} to the cluster rest frame,
is $1354^{+176}_{-118}$ km s$^{-1}$. 

This galaxy distribution has been studied using the techniques described 
by \scite{hobbs97a}, using the Jeans equation to relate the spatial and
velocity distribution of the galaxies to the gravitating mass profile for
the cluster. The aim of such an analysis is to determine the radial
behaviour of the anisotropy in the galaxy velocity distribution, since no
information about galaxy orbits is otherwise available. Since the radial
distributions of galaxy density and velocity are projected along the line
of sight, the analysis requires an extrapolation to large radius of the
X-ray determined mass profile, the galaxy velocity dispersion profile and
the galaxy surface density profile. Extrapolating the galaxy profiles is
particularly uncertain because data only extend out to $\sim2\arcmin$
($\sim500$kpc) and include no information (except in projection)
regarding the behaviour of these profiles beyond this region.

The anisotropy is studied through $\beta$, the anisotropy parameter,
which is 1 in the case of purely radial orbits, 0 for an isotropic
velocity distribution and increasingly negative as the orbits become
predominantly circular, with the limiting case of $-\infty$ for purely
circular orbits. It is unlikely, however, that the full range of allowed
values of $\beta$ is covered in any cluster. On the basis of numerical
simulations, \scite{yepes92a} found that over the lifetime of most
clusters, anisotropies corresponding to a $\beta$ more negative than -1.5
are unlikely to have had time to develop.

For the galaxy surface density profile, a standard modified-Hubble
profile ($\Sigma=\Sigma_0 \left[ 1+(r_p/r_c)^2 \right]^{-1}$), with a
canonical core radius of 250 kpc, has been used. The same
parameterisation has been adopted by \scite{natar96a}, who carried out an
analysis rather similar to that presented here. The line of sight
velocity distribution has been fitted by a linear ramp model using a
maximum likelihood method. The best-fit has a central velocity dispersion
of $1478^{+345}_{-321}$ km s$^{-1}$, while the gradient, although
consistent with zero, is such that the dispersion falls with radius at a
rate of $-114^{+257}_{-240}$ km s$^{-1}$ arcmin$^{-1}$.

\begin{figure*}
 \vspace*{9.5cm}
 \includegraphics{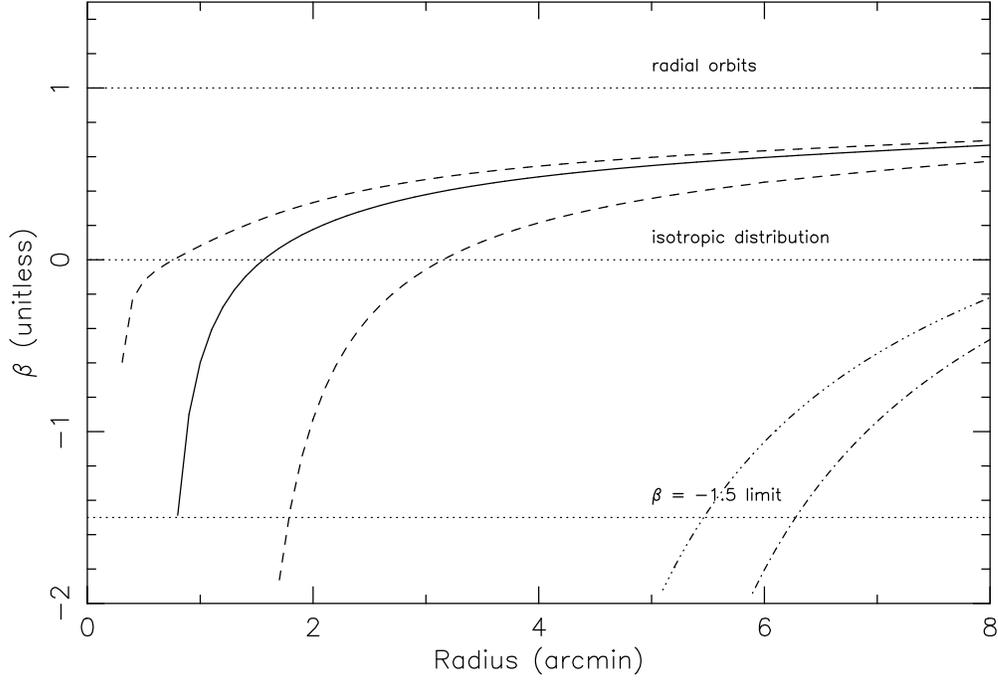}
 \caption[Derived galaxy anisotropy profiles for A2218]{\small 
 The anisotropy parameter for the two mass models discussed in the text is
 plotted as a function of radius. The anisotropy derived using the MMMC
 mass profile is indicated by the solid line, together with its 90\%
 uncertainty envelope (dashed lines). The MMM (dash-dot line) is shown
 together with its upper error bound (dash-3dot line) only, since the
 lower bound is unphysical at all radii. The MMMC is acceptable at
 r$\gtapp1\arcmin$, while the MMM is only allowed at radii beyond the
 observed edge of the optical data. Both models predict predominantly
 radial orbits at large radius. The $\beta=-1.5$ limit refers to the
 constraint from simulations (referred to in the text) that anisotropies
 more negative than $-1.5$ should not have had time to develop.}
\label{fig:beta_fig}
\end{figure*}

The results of the analysis are shown in Fig~\ref{fig:beta_fig}, in which
the calculated anisotropy parameter profiles for both of the most
interesting X-ray mass models (the MMM and MMMC) are shown. Although the
shape of the MMMC anisotropy profile is in detail different from those
presented by \scite{natar96a}, the conclusion is the same - there is a
divergence in the degree of anisotropy in the core of A2218. This is a
result of the high central line of sight velocity dispersion being
inconsistent with the mass provided by either the MMM or MMMC.

In our anisotropy profiles, the anisotropy plummets to $-\infty$ at a
non-zero radius (within the central $\sim1\arcmin$ in the case of the
MMMC model) and within this radius the solution is unphysical, requiring
an imaginary velocity dispersion. There are two possible reasons for
this. Firstly, in comparison with other clusters, A2218 has an unusually
large central velocity dispersion. Numerical simulations \cite{schind93a}
have shown that the velocity dispersion can increase by up to a factor of
2 during a merger event. This occurs during the violent relaxation phase
of the merger. At this time, the assumptions underlying any analysis
based upon the Jeans equation are invalid. This provides one way in which
the unphysical behaviour of the anisotropy parameter can be understood.

Secondly, the unphysical values of $\beta$ may be indicating that the
spherically symmetric model and assumptions that our solution uses may be
in error. This is supported by several sets of independent evidence which
suggest that the cluster core is disturbed, on the scale of
$\sim1\arcmin$. \scite{kneib95a} find that the strong lensing data
require a bimodal potential, \scite{markev97a} uses high-resolution X-ray
data to indicate that the cluster strongly deviates from spherical
symmetry in the core and recently \scite{girardi97a} showed that the
combined galaxy spatial and redshift data indicate the presence of two
merging galaxy sub-clumps. Thus, there are good reasons to believe that
A2218 has recently undergone a merger event, upsetting the virial
equilibrium in the cluster core.

In the case of the MMM model, the solutions are unsatisfactory throughout
the region for which galaxy data are available, such that a more
widespread upheaval would be required to account for the high velocity
dispersion.

\section{Sunyaev-Zel'dovich effect}

The Sunyaev-Zel'dovich microwave decrement \cite{suny72a} results from
inverse-Compton scattering of cosmic background photons by electrons in the 
cluster gas. The magnitude of the effect depends upon the integral of the gas 
pressure along the line of sight and hence has a different dependence on gas 
density than the X-ray surface brightness. By combining an analysis of the 
X-ray emission with the observed SZ effect, it is possible to determine the 
distance of the cluster, and hence H$_{0}$. Recent measurements 
(\pcite{birk94a}; \pcite{jones93a}; \pcite{saund96a}) have been used to place 
constraints upon the Hubble constant in this manner. \scite{birk94a} measure 
the decrement in a linear strip across the cluster, deriving the 1D profile.
\scite{jones93a} and \scite{saund96a} work with 2D images of the decrement,
allowing construction of a parameterised model for the observed microwave
decrement.

Several different observations of the decrement for A2218 have been combined 
with X-ray data, assuming an isothermal plasma, to constrain H$_{0}$ 
(\pcite{silk78a}; \pcite{mchard90a}; \pcite{birk94a}; \pcite{jones95a}). The 
results obtained vary widely, with the most disparate estimates being 
24$^{+23}_{-10}$ km s$^{-1}$ Mpc$^{-1}$ \cite{mchard90a} and 65$^{+25}_{-25}$ 
km s$^{-1}$ Mpc$^{-1}$ \cite{birk94a}. 

\begin{figure*}
 \vspace*{9.5cm}
 \includegraphics{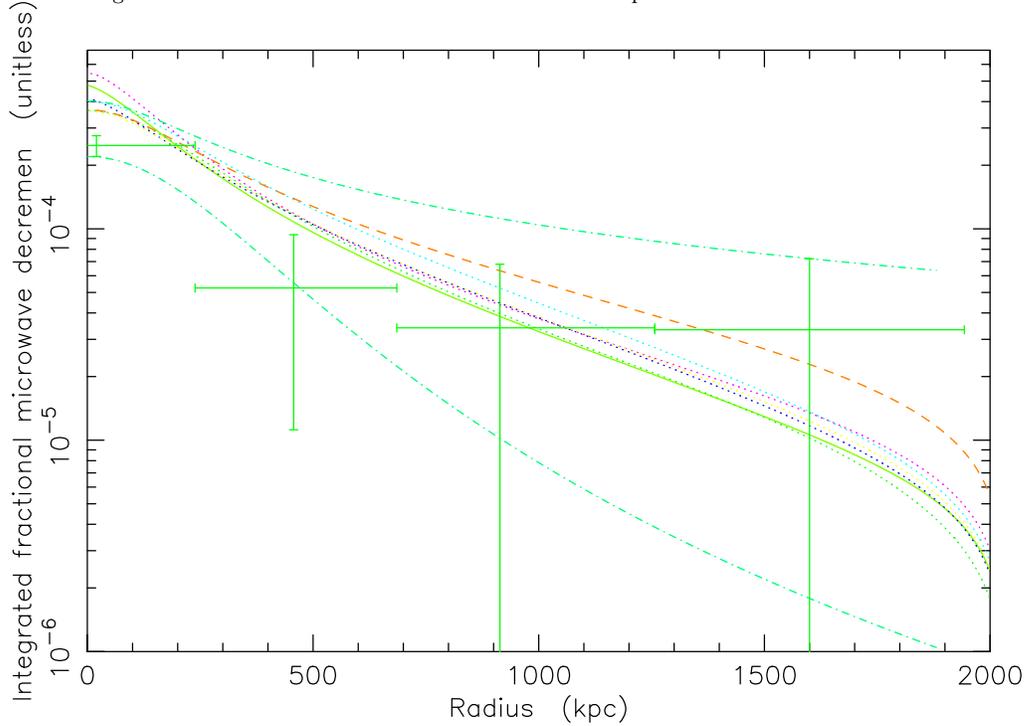}
 \caption[Comparison of SZ profiles]{\small
 Sunyaev-Zel'dovich decrement profiles predicted from the MMM (solid
 line) and best-fit cluster models (dotted lines), together with the
 decrement which occurs when an isothermal model is used (dashed line).
 The observed results of \scite{birk94a} (crosses) and \scite{jones93a}
 (dash-dot $1\sigma$ error envelope) are overlaid. A Hubble constant of
 50 km s$^{-1}$ Mpc$^{-1}$ is assumed.} \label{fig:sz_pred}
\end{figure*}

In Fig~\ref{fig:sz_pred} the predictions of our X-ray analysis are
compared with the measurements of \scite{birk94a} and the allowed
envelope of
\scite{jones93a}, assuming H$_0$=50 km s$^{-1}$ Mpc$^{-1}$. The typical
statistical uncertainty for a single model is similar to the scatter between 
the best-fit model predictions; both of these are small compared to the SZ 
error envelope.

At radii greater than 100 kpc, the observed and predicted decrements are
in excellent agreement. However, at smaller radii, the decrement observed
by \scite{birk94a} lies significantly below the X-ray model predicted
profiles. Limitations inherent to beam-switching single-dish measurements
are that they are prone to baseline errors and beam dilution. Thus our
analysis is not based upon the \scite{birk94a} data, but instead makes
use of their results only to allow a comparison with earlier studies.

These problems are not shared by the observation of \scite{jones93a},
whose calculated envelope encompasses both the \scite{birk94a}
observations and the majority of the predicted profiles for H$_0$=50 km
s$^{-1}$ Mpc$^{-1}$, with the MMM being an important exception. This
discrepancy occurs because the MMM requires a steep gas temperature
gradient at small radii (see Fig~\ref{fig:temp_fits}), resulting in a
high prediction for the central decrement. If the gas is assumed to be
isothermal (which is not statistically allowed by our X-ray data) the
predicted SZ decrement at large radius is significantly greater than that
from the best-fit ASCA models (see Fig~\ref{fig:sz_pred}). However, the
isothermal model cannot be discriminated against on this basis as it
remains consistent with the large SZ error envelope.

If an alternative value of H$_{0}$ is assumed, the predicted decrements
obtained from the best-fit X-ray models can be varied to achieve
consistency with the SZ observations. The dependency is such that for a
given X-ray surface brightness, the predicted SZ decrement varies as
H$_{0}^{-0.5}$. Hence, assuming a higher Hubble constant will lead to a
lower X-ray predicted decrement. The 2D SZ observation of
\scite{jones93a} and \scite{jones95a} is ideal for such a comparison
because, first, it avoids the uncertainty inherent in the \scite{birk94a}
analysis and, second, upper and lower bounds to the allowed decrement are
derived. However, it should be noted that these decrements have been
analytically parameterised to allow a fit to the mosaiced SZ image and
hence include a degree of model dependence \cite{jones95a}. With this
caveat in mind, it is possible to determine the range of H$_{0}$ allowed
by the SZ and X-ray results.

Under the requirement that at least one of the ASCA best-fit models must
be consistent with the results of \scite{jones93a}, we find that H$_{0}$
can be limited to the very conservative range 37-230 km s$^{-1}$
Mpc$^{-1}$. The lower bound is obtained by determining what value of
H$_{0}$ is required to make the SZ prediction from the LTF model (which
predicts the lowest central decrement of any of the models) equal to the
\scite{jones93a} upper bound. The upper bound, which is clearly ruled out
by other H$_{0}$ determinations, is obtained by determining the value of
H$_{0}$ which would decrease the SZ prediction from the TTF model (which
provides the highest central decrement of any of the models) such that it
matches the \scite{jones93a} lower bound. Using a value of H$_{0}$
outside these bounds results in all of the ASCA best-fit models becoming
inconsistent with the SZ observations.

The main conclusion of this analysis is that only weak constraints on the 
Hubble constant can be obtained, with even the very low \scite{mchard90a} 
result being allowed within its errors. 

However, this determination neglects an important additional constraint,
namely the gravitating masses extracted from lensing analysis. If we
consider only the MMM and MMMC, which have been shown to be the preferred
models when strong lensing is taken into account, the allowed range of
H$_{0}$ is more tightly constrained.

The predicted central decrements from the MMM and MMMC (for
H$_{0}$=$50$km s$^{-1}$ Mpc$^{-1}$) are 4.8x10$^{-4}$ and 4.0x10$^{-4}$
respectively. These are compared to the \scite{jones93a} upper bound of
4.0x10$^{-4}$. To achieve consistency between these results, H$_{0}$ must
be $>62$km s$^{-1}$ Mpc$^{-1}$ for the MMM and $>50$km s$^{-1}$
Mpc$^{-1}$ for the MMMC. Lower values of the Hubble constant ensure that
the predicted decrements do not lie within the
\scite{jones93a} bounds.

If the constraint of isothermality (at the fitted temperature of
$7.9^{+0.7}_{-0.6}$keV) is applied, the predicted central decrement is
3.7x10$^{-4}$. When the value of H$_{0}$ is allowed to vary, the range
allowed by the \scite{jones93a} bounds is 30-105 km s$^{-1}$ Mpc$^{-1}$.
This is consistent with the value of H$_{0}=38^{+18}_{-16}$ km s$^{-1}$
Mpc$^{-1}$ derived by \scite{jones95a}.

In summary, the constraints currently available from SZ observations
combined with X-ray measurements are too weak to constrain the Hubble
constant to greater accuracy than 37-230 km s$^{-1}$ Mpc$^{-1}$. This
range accommodates (within errors) the previously determined values of
24$^{+23}_{-10}$ km s$^{-1}$ Mpc$^{-1}$ \cite{mchard90a},
65$^{+25}_{-25}$ km s$^{-1}$ Mpc$^{-1}$ \cite{birk94a} and
38$^{+18}_{-16}$ km s$^{-1}$ Mpc$^{-1}$ \cite{jones95a}. However, when
lensing observations are introduced (which favour the MMMC model) the
Hubble constant is required to be greater than 50 km s$^{-1}$ Mpc$^{-1}$.
This result highlights the importance of using both non-isothermal gas
models and an approach which incorporates constraints in addition to
those provided by the X-ray and SZ observations alone.

\section{Discussion}

We have analysed ROSAT PSPC and ASCA GIS X-ray data, fitting spherically 
symmetric emission models to allow the extraction of cluster properties. The 
use of ASCA data allows the gas temperature structure to be discerned, a 
significant advance over earlier instruments. The analysis procedure utilises 
this information by fitting a variety of parametric forms for gas density and 
gas temperature or gravitating mass. This avoids restricting the unknown 
temperature profile to a single form.

Since the analysis presented here removes the constraint of isothermality, it 
is important to understand the effect that this assumption has. When 
isothermality is applied to Equation~\ref{eq:xray_hydro}, it degenerates to:

\begin{equation}
M_{grav}(r) = - {\frac {kT_{gas} r}{G \mu m_{p}}} \left[ {\frac
                {d \ln \rho_{gas}(r)}{d \ln r}} \right]
\end{equation}

Hence the {\it shape} of the mass profile is constrained by the gas
density profile alone. Since the latter is commonly taken to follow a
King model, the gravitating mass distribution is forced to take the form
of an isothermal sphere, with $\rho(r)\propto r$ at large radius and
flattening within a region determined by the gas core radius.

The main results are:

1) Comparison with strong lensing indicates that the previously reported
discrepancy is not present when the MMM is used. A consequence of the
extra mass required within the critical radius is a high central gas
temperature, which may be related to the disturbed nature of the core.
So, by relaxing the assumption of isothermality and using an appropriate
parameterisation for the dark matter distribution (one which includes a
mass cusp), the observed X-ray and strong lensing data can become
consistent.

2) The suggestion, by \scite{makin96a}, that the cD contributes a
significant fraction of the cluster mass at small radius has been tested.
The maximum central mass which can be added to the MMM, before the Cash
statistic shifts beyond the 95\% error level, is 1.7x10$^{12}$
M$_{\odot}$ (within 100kpc). Thus the massive galaxy required by
\scite{makin96a} is ruled out, if the underlying DM distribution follows
the form of the MMM model. The effect of adding the maximum allowed
central galaxy mass is to extend the DM distribution. Because of this,
the MMMC is found to be more consistent than the MMM with the outer two
data points extracted from the \scite{kneib95a} strong lensing analysis.

3) At larger radii, 200-1000 kpc, the projected mass profiles of the MMM
and MMMC are consistent with the weak lensing results of
\scite{squire96a}. When the slope of the mass profile is considered, the
MMMC becomes the favoured model, as the MMM distribution is considerably
flatter than the trend of the weak lensing data. It is important to
recognise that the weak lensing analysis provides only a lower limit to
the actual mass distribution (since cluster mass is known to reside in
the weak lensing control annulus). Hence, although the MMMC provides more
mass than the weak lensing results currently allow, this discrepancy may
be resolved when statistical lensing of the background galaxy population
is observed to greater radii (so that a control annulus beyond the edge
of the cluster can be used).

4) The steep central rise in temperature required by the MMM leads to a
corresponding increase in entropy. Under these conditions the gas is
likely to be convectively unstable within $\sim 120$kpc. This is either a
reflection of the true physical state of the ICM or a model-dependent
effect, related to the parametric forms assumed for the gas density and
temperature. Evidence for the former is provided by \scite{markev97a},
whose HRI analysis indicates the existence of central substructure,
perhaps as a result of merging activity. If this is the case, the gas is
likely to have been violently shock-heated, such that the assumption of
equilibrium within the cluster core is no longer secure. On the other
hand, the problem of convective instability is lessened with the MMMC,
due to the lower fitted central temperature. In addition, replacing the
King gas density profile with an NFW description can resolve the problem
entirely. This is a consequence of allowing the density to rise, rather
than flatten, in the cluster core.

5) Analysis of galaxy orbits provides another probe of the core region of
A2218. Using the derived anisotropy of the galaxy velocities, it is
possible to extract information about the gravitational potential and
possible presence of substructure. Beyond the central 1$\arcmin$ region,
the velocity dispersion data are consistent with the MMMC. At smaller
radii, a physically reasonable solution for the anisotropy parameter
cannot be attained with any of the derived X-ray mass distributions. This
suggests that the central velocity dispersion has been raised to its high
observed value during the violent relaxation phase of a merger event.
Thus, even though the galaxy data are too poorly sampled spatially to
indicate the presence of substructure, the velocity information supports
the analyses of \scite{kneib95a} and \scite{markev97a}.

6) Comparison with 2D Sunyaev-Zel'dovich observations indicates that the
Hubble constant is only weakly constrained, to a range of 37-230 km
s$^{-1}$ Mpc$^{-1}$. This is due to the high level of uncertainty which
occurs when errors in the SZ and X-ray data are combined. More
restrictive constraints on the value of H$_{0}$ can be obtained if
additional information is used, such as lensing observations. When the
MMM and MMMC are combined with the SZ data, low values of the Hubble
constant are ruled out and we find that H$_{0}>50$ km s$^{-1}$
Mpc$^{-1}$. This result conflicts with the previously determined value of
24$^{+23}_{-10}$ km s$^{-1}$ Mpc$^{-1}$ \cite{mchard90a}, but agrees with
the later estimate of 65$^{+25}_{-25}$ km s$^{-1}$ Mpc$^{-1}$
\cite{birk94a}. The differences between these results, and that derived
here, are dominated by several factors. The first, and most important, of
these is that an unwarranted assumption has been made about the gas
temperature in these earlier studies - that it is isothermal. This biases
the analysis and also results in over-optimistic error estimates.
Secondly, \scite{mchard90a} and \scite{birk94a} both use (different) 1D
SZ measurements, which are prone to baseline uncertainties. Third, these
studies use the less accurate gas density information from Einstein,
rather than ROSAT.

Note, however, that these results are all based upon the commonly made
assumption that the X-ray and SZ observations can be made consistent
purely by manipulating the value of H$_{0}$. If the cluster gas, at small
radius, is not in hydrostatic equilibrium, this methodology may be in
error.

\section{Conclusions}

Combining the above results, it appears that A2218 consists of two
distinct regions. At small radii ($\leq1\arcmin$), the X-ray morphology
is extremely disturbed \cite{markev97a} with both the strong lensing
\cite{kneib95a} and galaxy data \cite{girardi97a} indicating the presence
of bimodal structure. An X-ray model (the MMMC) consistent with the
strong lensing data requires the gas temperature to rise steeply within
the core, possibly at such a high rate that the gas is convectively
unstable. When this model is combined with the observed galaxy data, a
physical solution cannot be obtained for galaxy orbits within the core.
Taken together, these results suggest that A2218 has recently undergone a
merger, shock-heating the gas and disturbing the equilibrium of the
components within the cluster. This explains the lack of a cooling flow
in the system.

Outside the core, our results show that the data can be consistently
interpreted on the basis of a cluster in equilibrium. The gas temperature
falls to $<10$ keV, consistent with the observed galaxy velocity
dispersion, and the gas entropy increases with radius. The MMMC provides
a mass profile consistent with both the outer two strong lensing points
and the weak lensing mass profile (with the proviso that the latter is
likely to represent an underestimate of the true mass profile). When the
MMMC is combined with the galaxy velocity data, a physical solution for
the galaxy orbits is recovered. The SZ data, which are sensitive to the
gas pressure at large radius, are consistent with the MMMC when
H$_{0}\geq$50 km s$^{-1}$ Mpc$^{-1}$.

Thus, while it is premature to regard a model such as the MMMC as a
complete description of the physical structure of A2218, it does appear
to explain all the data available beyond the disturbed core. To probe
further, more detailed X-ray and weak lensing observations are required.

One of the principal conclusions to be drawn from the present analysis is
that it is dangerous to assume an isothermal ICM without supporting
evidence. This can lead to biased conclusions and underestimated errors.

\section*{Acknowledgments}

We would like to thank Gordon Squires for providing us with his weak
lensing mass profiles, Mike Jones for his information on interferometric
Sunyaev-Zel'dovich analysis and Paul Nulsen \& Priya Natarajan for
helpful discussions. In addition, we wish to the referee, Hans
Boehringer, for valuable suggestions which improved the paper.

DBC and ISH acknowledge the financial support of the UK Particle Physics and
Astronomy Research Council (PPARC). Computational work was done on the
Birmingham node of the PPARC funded Starlink network.

\bibliographystyle{mn}
\bibliography{2218_paper}

\bsp

\label{lastpage}

\end{document}